\documentclass[12pt]{iopart}
\pdfoutput=1 

\expandafter\let\csname equation*\endcsname\relax
\expandafter\let\csname endequation*\endcsname\relax

\usepackage{cite} %to make contiguous citations to appear like [1-3]

\usepackage{amsmath,amssymb}
\usepackage{mathtools}
\usepackage{siunitx}

\DeclarePairedDelimiterX\braket[2]{\langle}{\rangle}{#1\,\delimsize\vert\,\mathopen{}#2}
\usepackage{lineno}

\usepackage{color}
\definecolor{intreview}{RGB}{0, 0, 0}
\definecolor{review}{RGB}{200, 0, 0}
\definecolor{ready}{RGB}{0, 128, 0}

\usepackage[usenames,dvipsnames]{xcolor}
\definecolor{linkcolor}{rgb}{0.0,0.3,0.5}

\usepackage[hidelinks]{hyperref}
\hypersetup{colorlinks=true, linkcolor=linkcolor, linktocpage=true,  citecolor=linkcolor, urlcolor=linkcolor}
\usepackage{multirow} 
\usepackage{subfig}
\usepackage{graphicx}
\usepackage{caption}
\usepackage{siunitx}
\usepackage{orcidlink}

\newcommand{\keywords}[1]{\small \textbf{Keywords:} #1}

\newcommand{\vect}[1]{\boldsymbol{#1}}
\usepackage[symbol]{footmisc}

\usepackage{fontawesome5}  % For GitHub logo
\definecolor{kenncolor}{HTML}{CF4587}

\begin{document}
%\linenumbers

%==============================================
%========== CONVENCTIONS =======================
%==============================================
%FONTS PER NOMI DETECTOR
%ACRONIMI CONCESSI:
%CONVENZIONE EQUAZIONI

%STRUCTURE
%INTRO
%OVERLAPPING
%METHODS
% - intro to the pipeline
% - KENN description + transformer encoders
% - flows + HYPERION
%SIMULATIONS & TRAINING 
%RESULTS & DISCUSSION
%CONCLUSIONS

% ------------------- Frontmatter ------------------- %
%\title[Transformers]{I'm just KENN: a Transformer Encoder for parameter estimation of overlapping signals in the Einstein Telescope}
\title[Transformers for gravitational wave overlapping signals]{Can Transformers help us perform parameter estimation of overlapping signals in gravitational wave detectors?}

\author{Lucia Papalini$^{1,2, \text{\footnotemark[1]}}$\footnotetext[1]{\href{mailto:lucia.papalini@phd.unipi.it}{lucia.papalini@phd.unipi.it}}$\,$\orcidlink{0000-0002-5219-0454}, Federico De Santi$^{3,4,\text{\footnotemark}}$\footnotetext{\href{mailto:lf.desanti@campus.unimib.it}{f.desanti@campus.unimib.it}}$\,$\orcidlink{0009-0000-2445-5729}, Massimiliano Razzano$^{1,2}$$\,$\orcidlink{0000-0003-4825-1629}, Ik Siong Heng$^{5}$$\,$\orcidlink{0000-0002-1977-0019}, Elena Cuoco$^{6,7}$$\,$\orcidlink{0000-0002-6528-3449} }

\address{$^1$Dipartimento di Fisica ``Enrico Fermi", Universit\`a di Pisa, Largo B. Pontecorvo 3, I-56127 Pisa, Italy}
%\address{$^2$Istituto Nazionale di Fisica Nucleare, Sezione di Pisa, I-56127 Pisa, Italy}
\address{$^2$INFN, Sezione di Pisa, Largo B. Pontecorvo 3, I-56127 Pisa, Italy}
\address{$^3$Dipartimento di Fisica ``G. Occhialini'', 
Universit\`a degli Studi di Milano-Bicocca, Piazza della Scienza 3, 20126 Milano, Italy}
\address{$^4$INFN, Sezione di Milano-Bicocca, 
Piazza della Scienza 3, 20126 Milano, Italy}
\address{$^5$School of Physics and Astronomy, University of Glasgow, G12 8QQ, United Kingdom}
\address{$^6$Physics and Astronomy Department (DIFA), Alma Mater Studiorum - Università di Bologna, Italy} 
\address{$^7$INFN, Sezione di Bologna Viale C. Berti Pichat 6/2 - 40126 Bologna, Italy}

\begin{abstract}
Overlapping signals represent one of the major data analysis challenges in next-generation gravitational wave detectors. We leverage Transformers and Normalizing Flows, state-of-the-art machine learning algorithms, to address the parameter estimation of overlapping binary black hole mergers in the Einstein Telescope (ET).
Our proposed model combines a Transformer-based ``Knowledge Extractor Neural Network" (\textsc{KENN} \href{https://github.com/luciapapalini/kenn-gw-transformer.git}{\textcolor{kenncolor}{\faGithub}}) with a Normalizing Flow (\textsc{HYPERION} \href{https://github.com/fdesanti/HYPERION}{\faGithub}) to perform rapid and unbiased inference over multiple overlapping black hole binary events. The choice of architecture leverages the strength of Transformers in capturing complex and long-range temporal structures in the strain time series data, while Normalizing Flows provide a powerful framework to sample posterior distributions. We demonstrate the effectiveness and robustness of our model over simulated gravitational wave signals, showing that it maintains the same level of accuracy regardless of the correlation level in the data. Moreover our model provides estimates of chirp mass and coalescence times within $\lesssim 10-20\%$ from the true simulated value. The results obtained are promising and show how this approach  might represent a first step toward a deep-learning based inference pipeline for ET and other future gravitational wave detectors.
\end{abstract}


\submitto{\CQG}
\maketitle
%-----------------------INTRODUCTION-----------------------

\section{Introduction}
Gravitational wave science has advanced significantly since the first detection of a Binary Black Hole (BBH) signal by the LIGO-Virgo-KAGRA (LVK) Collaboration a decade ago \cite{GW150914}. The increasing detection rate \cite{cumulative_detection_rate}, allowed by substantial instrumental upgrades over the different observing runs, has stimulated the development of new data analysis techniques and pipelines.\\
In roughly a decade from now,  next-generation detectors such as the Einstein Telescope (ET) \cite{ET}, Cosmic Explorer (CE) \cite{Cosmic_Explorer} and LISA \cite{LISA} are expected to become operative. The unprecedented sensitivity of these detectors will pose several challenges to current data analysis approaches.  Among these, the issue of overlapping signals from coalescing binary systems stand out as critical for ET and CE and, to some extent, in LISA. For instance, the Einstein Telescope will observe $\mathcal{O}(10^{5}\text{ events}/\text{yr})$ \cite{Iacovelli:2022bbs}, each one with longer durations compared to current LVK detections \cite{ET_science_case_different_designs} due to the better low-frequency sensitivity. This will lead to overlapping signals that will make unpractical the use of single-event parameter estimation (PE) pipelines. In particular, addressing the challenge of a fast and unbiased PE \cite{PE_challenges_3G, Pizzati_et_al_overlap, Overlapping_samajdar} is essential to accurately resolve single sources either to perform GR tests \cite{impact_overlapping_GR_tests} or ensure precise subtraction \cite{cbc_subtraction_for_sgwb_2, cbc_subtraction_for_sgwb}, which is critical to detect and study the stochastic gravitational wave background \cite{overlapping_cosmological_sgwb, notching_cosmological_sgwb}, or even perform population studies \cite{ET_blue_book}.\\
The nature of these challenges is evolving, as it is the way we approach and address them. A notable development is the growing role of machine learning (ML) \cite{Cuoco:2024cdk, ML_physics_GW}. Over the years, machine learning has been applied to a wide range of tasks \cite{Cuoco:2024cdk}, including detection \cite{Nousi_AresGW, Bini_ML_cWB, ML_detection_SNe, ML_detection_long_transients}, parameter estimation \cite{Dingo, Nessai, Gabbard_Vitamin}, detector characterization \cite{Razzano_gwitchhunters, Gravity_Spy, denoising_ML} and control \cite{Interferometer_Control_ML}. Machine learning often provides faster solutions compared to traditional methods. Among these, its contributions to parameter estimation, in the framework of simulation-based inference (SBI), has been particularly significant. In this context, models such as Normalizing Flows have demonstrated to be exceptionally promising and successful \cite{Dingo, DeSanti_CE_HYPERION}.\\
In the recent years,\textit{ Transformers} \cite{vaswani2017attention} have gained significant attention within the ML field for their effectiveness in processing large sequential data for a variety of tasks such as natural language processing \cite{kenton2019bert, achiam2023gpt} and audio modeling \cite{radford2023robust}.\\
In this work, we present a first  application of Transformers to gravitational wave data analysis of overlapping CBC signals in the Einstein Telescope, with the possibility of extending this approach to other next-generation detectors.

The paper is organized as follows. In Sec. \ref{sec:overlapping_sig} we discuss motivations and theoretical background of this work, illustrating the challenges posed by signal overlap in ET. In Sec. \ref{sec: methods} we present in detail our implementation of Transformer-based architecture and its integration with Normalizing Flows, as well as the simulation and training processes. In Sec. \ref{sec: results and discussion} we discuss the results and performance of our model. In Sec. \ref{sec: conclusion} we summarize the implications and potential future developments of this approach.

%-----------------------MOTIVATION-------------------------
\section{Overlapping Signals}
\label{sec:overlapping_sig}

\begin{figure}[t]
    \centering
    \includegraphics[width=0.85\linewidth]{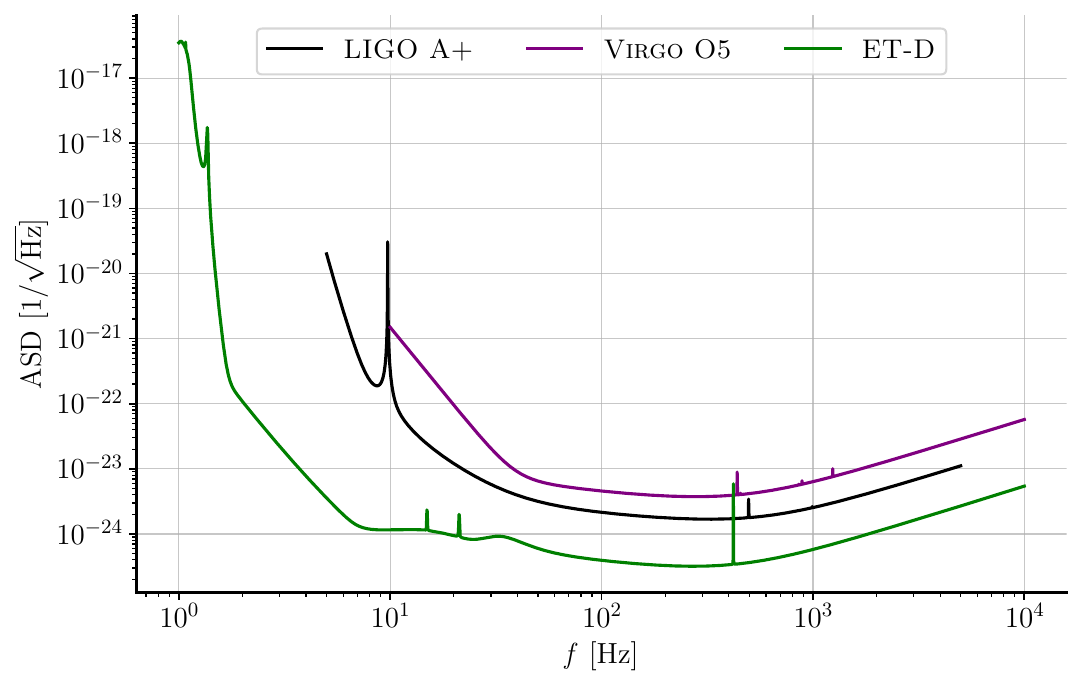}
    \caption{Design sensitivity curves comparison of \textsc{LIGO A+} \cite{LIGO_A+_curve}, \textsc{Virgo O5} \cite{LIGO_A+_curve} and \textsc{ET} \cite{ET_curve}.} 
    \label{fig: ET sensitivity curve}
\end{figure}

%Due to its technological advances, 
Einstein Telescope is expected to reach a design sensitivity at least one order of magnitude greater than current detectors in a frequency band region spanning from $\sim \SI{5}{\hertz}$ to a few $\SI{}{\kilo \hertz}$: see Fig. \ref{fig: ET sensitivity curve} for a comparison of the sensitivities of ET, LIGO A+ and Virgo O5. \\
 Thanks to its sensitivity  ET will observe the coalescence of stellar-mass BBH coalescences up to redshift $z\sim 20$ and binary neutron stars (BNS) up to $z\sim 2-3 $, therefore boosting the rate of events up to $ \mathcal{R}_{\text{BBH}} \sim 10^5-10^6$ events/yr and $ \mathcal{R}_{\text{BNS}}\sim 7\times10^4$ events/yr \cite{ET}.  
Furthermore, increasing the sensitivity band towards the low frequencies implies the possibility of observing more cycles of the inspiral phase before the merger. Indeed, for equal-mass non spinning binaries, the time spent into the detector's sensitivity band is given by \cite{Overlapping_samajdar}

\begin{equation}\label{eq: signal length}
    \tau = 2.18\left(\frac{1.21 M_\odot}{\mathcal{M}} \right)^{5/3} \left[\left( \frac{\SI{100}{\hertz}}{f_\text{low}} \right)^{8/3} -  \left( \frac{\SI{100}{\hertz}}{f_\text{high}} \right)^{8/3} \right] \SI{}{\second}
\end{equation}
with $\mathcal{M}$ being the chirp mass of the binary. 
In Eq.\eqref{eq: signal length} $f_\text{low}$ is the detector's lower sensitivity bound and $f_{high}$ can be considered the frequency at the Innermost Stable Circular Orbit (ISCO)
\begin{equation}\label{eq: ISCO frequency}
    f_\text{ISCO} = \frac{1}{6\pi\sqrt{6}} \frac{c^3}{GM} \simeq \SI{2.2}{\kilo\hertz} \left( \frac{M_\odot}{M} \right)
\end{equation}

\noindent with $M$ the total mass of the binary.\\
According to Eq. \eqref{eq: signal length}, such gain in the $f_\text{low}$ implies that a given event can spend $\sim200$ more time in the sensitivity band$^\text{\footnotemark}$\footnotetext{Assuming a low frequency cutoff at $\SI{20}{\hertz}$ and $\SI{5}{\hertz}$ for LIGO and ET respectively.}. For instance, an event with $\mathcal{M}\sim 30 M_\odot$ like GW150914 \cite{GW150914} would have stayed in the sensitivity band for over $\sim3$ minutes instead of $\SI{0.75}{\second}$. Similarly, a BNS signal like GW170817 with $\mathcal{M}\sim 1.19$ \cite{gw170817} would have been observable around 2 hours before the merger instead of just $\sim \SI{160}{\second}$. 

The combination of the high event rate and the larger signal duration inevitably results in the presence of overlapping signals, raising two competitive challenges: computational cost and biases in the parameter estimation.
These have been studied in several works \cite{Overlapping_samajdar, Janquart_joint_PE, Pizzati_et_al_overlap, PE_overlapping_signals_2G, Antonelli_biases_overlapping, Himemoto_overlapping_cgwb, Wang_PE_bias_analusis_with_FM, Source_confusion_BNS_minimal_overlapping_PE}. More precisely, \cite{Overlapping_samajdar} estimates the number of overlapped signals within a year of observation adopting astrophysically motivated merger rate densities $\mathcal{R}_\text{GW}(z)$ \cite{merger_rate_density_used_in_samajdar_1, merger_rate_density_used_in_samajdar_2} for both BBH and BNS. Depending on merger rates uncertainties$^\text{\footnotemark}$\footnotetext{due to the uncertainties on the primary mass distribution $p(m_1)$ \cite{gwtc3_population}}, they estimate that $11^{+4}_{-7}$ BBH mergers will happen within \SI{2}{\second} apart on average.\\ 
As highlited in \cite{Pizzati_et_al_overlap, PE_overlapping_signals_2G}, significant biases can occur when recovering parameters of signals with close merger times. In \cite{Pizzati_et_al_overlap} a Fisher Matrix formalism is used to address and quantify the biases that arise at $2-3\sigma$ level when the difference in merging time $|\tau| \lesssim \SI{1}{\second}$.
A systematic study considering different parameter estimation methodologies for the analysis of just two overlapping signals is performed in \cite{Janquart_joint_PE} where joint parameter estimation and hierarchical subtraction are the methods being tested. 
Although computationally cheaper, this last method is more prone to biases. On the other hand, joint analysis has a much higher computational burden, requiring up to months. These challenges highlight the need for new and faster analysis algorithms. \\
%Here we present a novel approach based on Deep Learning exploiting state of the art architectures: Transformers \cite{vaswani2017attention} and Normalizing Flows \cite{papamakarios_flows}. \\

\section{Methods}\label{sec: methods}

To address the challenge of multiple overlapping signals' parameter estimation, we propose a novel method combining a Transformer-based \cite{vaswani2017attention} architecture with a Normalizing Flow. We show the overall scheme in Fig. \ref{fig: kenn&model}.
Our model is designed to analyze data from the three ET data streams (channels) and estimate parameters for three injected BBH signals, randomly overlapped. To train it, we implemented a custom Dataset Generator that dynamically creates new simulations during training, providing a varied training set.

\subsection{\textsc{KENN}: the Transformer Encoder}\label{sec: KENN-Transformers}

The Transformer component in our model is designed to extract relevant features from input data by representing them in the embedding space. Hereafter, we will refer to it as the Knowledge Extractor Neural Network (\textsc{KENN}).

The Transformer \cite{vaswani2017attention} is an architecture originally designed as a language translation model, and it introduced a mechanism designed to overcome sequential data models limitations. In fact, traditional sequential models like Recurrent Neural Networks (RNN) \cite{cho2014learning}, Long Short-Term Memory (LSTM) \cite{hochreiter1997long}, and Gated Recurrent Units (GRUs) \cite{dey2017gate} processed data recursively, meaning that each hidden state of the network $h_t$ depends on the previously calculated state $h_{t-1}$. This kind of recursive dependency can lead to an effect known as vanishing gradient \cite{hochreiter2001gradient}, that hiders the capability of a neural network to capture long-range dependencies in the sequence.
In this context, the key aspect that has made the Transformer so versatile and popular is its ability to capture long-range dependencies while maintaining parallelizable computation.

This architecture typically consists of two main components: an encoder and a decoder. The encoder processes the entire input sequence simultaneously, transforming it into a more refined representation that captures structural relationships and correlations among its elements. The decoder, on the other hand, generates the output sequence step by step, taking as input both the encoded representation and the previously generated parts to produce coherent and contextually relevant outputs. Neither component operates directly on the original time series. Instead, they operate on a representation that has been translated into a trained embedding space, a high-dimensional vector space where segments of the original sequence are mapped by an embedding layer to optimally encode their information. Its dimensionality, denoted as $d_\text{model}$, defines the size of these vector representations. The segment of data that is encoded into a single vector is typically called token. In language processing, it can represent parts of words or phrases, while in time series data, it corresponds to discrete chunks of a sequence.\\ Both the encoder and the decoder consist of stacked identical layers, each built around similar core mechanisms. At their heart lies the \textit{multi-head self-attention mechanism}, which plays the central role, followed by feed-forward layers. These components are connected through residual connections and layer normalization, ensuring the preservation of information from earlier stages.

\begin{figure}[!t]
    \centering
    \includegraphics[width=\linewidth]{ 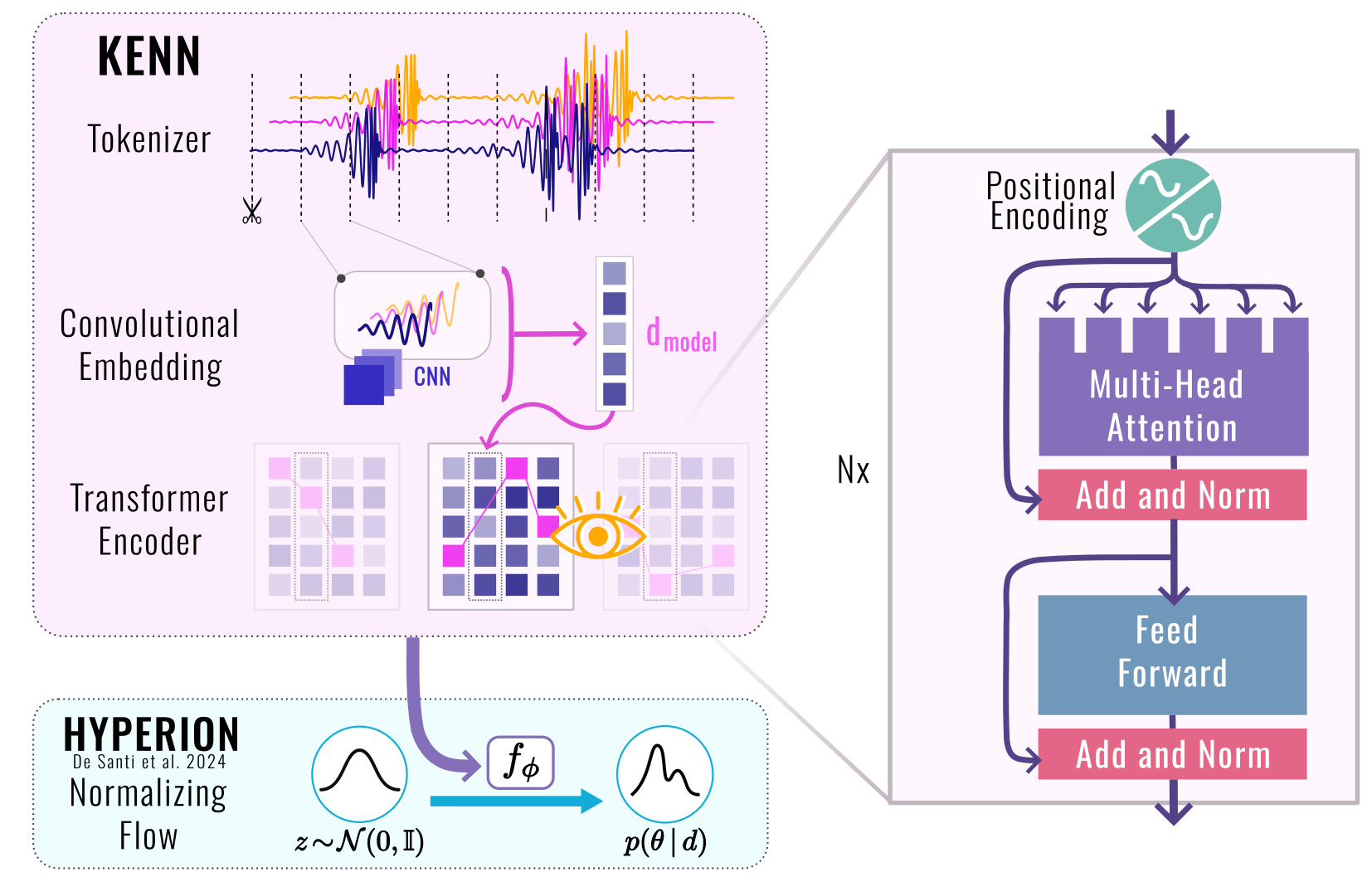}
    \caption{Schematic overview of our model. The pink box shows the KENN architecture. The three ET channels are first split into chunks by the Tokenizer. These chunks are then passed to the Convolutional Embedding module, which maps them into an embedding space of dimension $d_\text{model}$. Next, the Transformer Encoder, composed of $N=2$ identical layers, extracts the most relevant features from the data. The output is then passed to the Normalizing Flow architecture (\textsc{HYPERION}) in the blue box, which reconstructs the posterior distribution of the sources' parameters.}
    \label{fig: kenn&model}
\end{figure}

Attention is a mechanism designed to capture the relationships between different elements of a sequence, independently from their distance in it. It is computed using the Scaled Dot-Product mechanism, which determines the relationship between different elements. It takes as input queries and keys of dimension $d_k$ and values of dimension $d_v$. Queries, keys, and values can be understood as components of an information retrieval system: each query represents a request for relevant information, keys serve as reference points that help identify the most relevant responses, and values contain the actual information to be retrieved. In the attention mechanism, the model learns to compare queries with keys to determine the importance of each value in the context of the given sequence. \\
The calculation of attention weights is done simultaneously on a set of queries, keys and values, arranged in matrices $Q$, $K$ and $V$. The result is the so-called Attention Matrix

\begin{equation}
    \text{Attention}(Q,K,V) = Softmax\Big(
    \frac{Q\cdot K^\top}{\sqrt{d_k}}\Big)V
\end{equation}

Where the Softmax function, $Softmax(x_i)=e^{x_i}/\sum^{n}_{j=1}e^{x_j}$, is used for making the weights a probability distribution. 

To enhance the performance of the attention mechanism, Multi-Head Attention extends the single attention mechanism by learning multiple attention representations in parallel. 
In particular, queries, keys and values are linearly projected from a dimension of $d_\text{model}$ to $d_k$, $d_k$ and $d_v$ dimensions to calculate $h$ parallel attention matrices. These different ``heads" are then concatenated, and the network learns different representations of the same sequence at the same time. The so-called \textit{Multi-Head Attention} mechanism is represented as follows

\begin{equation}
    \text{MultiHead}(Q,K,V) = Concat(h_1,..., h_h)W^O
\end{equation}

\noindent With the single head $h_i$ is defined as

    \begin{equation}
        h_i = \text{Attention}(QW^Q_i, KW^K_i, VW^V_i)
    \end{equation}
\noindent where the matrices %are made of learnable parameters 
$W_i^Q \in \mathbb{R}^{d_\text{model} \times d_k}$, 
$W_i^K \in \mathbb{R}^{d_\text{model} \times d_k}$, 
$W_i^V \in \mathbb{R}^{d_\text{model} \times d_v}$, 
and $W^O \in \mathbb{R}^{h d_v \times d_\text{model}}$ are made of learnable parameters.\\

%%%%%_____________________OUR MODEL_______________________%%%%%

Our implementation processes three-channel time series data using several neural network components like convolutional layers, multi-head attention, and positional encoding. A summary of all components can be seen in Fig. \ref{fig: kenn&model}. The Tokenizer module takes as input the three-channel strain time series and slices  them respectively into $L$ chunks of 0.5 seconds.
The Convolutional Embedding module consists of a CNN layer applied to each chunk to extract local features, followed by a linear transformation to project these features into an embedding space of dimension $d_\text{model}$. Therefore, the output of this part of the architecture is a matrix $L \times d_\text{model}$. 
This approach is supported by methodologies previously developed in other domains, such as audio and speech recognition models like Whisper \cite{radford2023robust}. Our Convolutional Embedding has 3 input channel and a single output channel, the kernel size is of 65 and the activation function is the ReLU function, as it is in all parts of the code. \\
Since, the information about the relative position of each encoded chunk is not retained by this procedure, we employ a Positional Encoding technique, as done in \cite{vaswani2017attention}.
%We use sine and cosine functions of varying frequencies as done in \cite{vaswani2017attention}, to ensure unique representations for each position, which are then summed with the input embeddings.
To represent the relative positions of the entries of each $k$-th ($0 \leq k < L/2$) column, we employ sine and cosine functions of varying frequencies

%Suppose we have an input sequence of length $L$, and we want to represent the position of the $k$-th object within this sequence, ranging between $0 \leq k < L/2$ . The positional encoding can be given by

\begin{equation}\label{eq: postional encoding}
    P(k, 2i) = \sin\left(\frac{k}{n^{{2i}/{d_\text{model}}}}\right)\:; \quad P(k, 2i + 1) = \cos\left(\frac{k}{n^{{2i}/{d_{\text{model}}}}}\right)
\end{equation}
where $n$ is a scalar set to $10^4$ \cite{vaswani2017attention} and $i$ is the index used for mapping to column indices $0 \leq i < d_\text{model}/2$, in both sine and cosine functions. 

The Transformer Encoder layer follows a standard architecture, incorporating multi-head attention and feed-forward networks.
The encoder layers also employ layer normalization and dropout to stabilize the training. In our case, the encoder is composed of 2 layers, with 32 attention heads, $d_\text{model} = 512$ and two layers with 2048 units each in the feed-forward networks. We employ dropout at multiple stages, including both attention and feed-forward layers. The dropout rate is set to 0.4 in the first part and 0.2 in the latter. %We adopt the \texttt{ReLU} activation function.\\ 

\subsection{Normalizing Flows}\label{sec: Normalizing Flows}
Normalizing Flows are a powerful class of Deep Generative Models \cite{DeepGenerativeModels} able to approximate complex probability distributions \cite{papamakarios_flows, RealNVP}. They can be employed in the SBI context to sample the posterior $p(\vect{\theta}_\text{gw}|\vect{d})$ \cite{Dingo, DeSanti_CE_HYPERION}. More in detail, the Flow learns a bijective map $f:\Theta_\text{gw}\longrightarrow \mathcal{Z}$ from the physical parameter space to a latent space in which samples are assumed to be normally distributed: $z\sim \mathcal{N}_{(\vect{0}, \mathbb{I})}$.

\begin{equation}
    p(\vect{\theta}_\text{gw}|\vect{d}) \simeq  q_\phi(\vect{\theta}_\text{gw}|\vect{d})= \mathcal{N}_{(\vect{0}, \vect{\mathbb{I}})}(f_\phi(\vect{\theta}_\text{gw}; \vect{d}))\left|  \det \mathcal{J}\left( \frac{\partial f_\phi(\vect{\theta}_\text{gw};\vect{d})}{\partial \vect{\theta}_\text{gw}}\right)\right|
\end{equation}
with $\mathcal{J}(\cdot)$ the Jacobian of $f_\phi$. The map $f_\phi$ is commonly parametrized with neural networks whose weights $\phi$ are optimized via stochastic gradient descend methods. The commonly adopted cost function is given by the Kullback-Leibler divergence $\mathrm{KL}[p||q_\phi]$ between the true posterior and the flow surrogate, which takes the following form \cite{papamakarios_flows}

\begin{equation}\label{eq: KL loss}
    \mathrm{KL}[p||q_\phi] = \mathbb{E}_{[p(\vect{d})] }\int d\vect{\theta}_\text{gw} \; p(\vect{\theta}_\text{gw}|\vect{d}) \log \left( \frac{p(\vect{\theta}_\text{gw}|\vect{d})}{q_\phi(\vect{\theta}_\text{gw}|\vect{d})} \right) \simeq - \frac{1}{N} \sum_{i=1}^N \log q_\phi(\vect{\theta}_\text{gw}|\vect{d})
\end{equation}
where the final expression results from a Monte Carlo approximation \cite{dingo_15D_toy, DeSanti_CE_HYPERION}.\\
Once the flow has been trained, the posterior is sampled by computing the inverse $f_\phi^{-1}$ on samples drawn from the latent base distribution.

%%%%%_____________________OUR MODEL_______________________%%%%%
The representation of data, learned by \textsc{KENN} is then fed to the flow to model the posterior distribution. For the Normalizing Flow we adopt the \textsc{HYPERION} sampler whose architecture is based on Affine Coupling Transformation, as described in more detail in \cite{DeSanti_CE_HYPERION}. The flow model has 32 coupling layers. \textsc{HYPERION} and \textsc{KENN} are jointly trained to minimize the cost function of Eq. \eqref{eq: KL loss}.

\subsection{Simulations}\label{sec: simulations}

%The Dataset Generator is a tool we built that enhances simulation efficiency by creating task-specific datasets.
%For this work, we fixed to three the number of overlapped signals to include in each time series sample. 
Our training dataset consist of pairs $\left(\{\vect{\theta}_\text{gw}\}_k, \vect{d}\right)$ where $\{\vect{\theta}_\text{gw}\}_k$ is the set of waveform parameters of $K=3$ overlapping BBHs and $\vect{d}$ the corresponding ET data stream in the time domain. For ET, we assume the triangular configuration and the Sardinia site location \cite{Iacovelli:2022bbs, Naticchioni:sos_enattos}. The simulations were run with a custom pipeline able to deliver new training sample online as the training runs. More in detail, data generation proceeds as follows.

\begin{enumerate}
    \item \textit{Simulation of template waveforms:} At the beginning of each epoch we draw 5000 samples from the \textit{intrinsic} parameters prior (Tab. \ref{tab: prior distributions}) to generate a template bank. We adopt the \texttt{IMRPhenomXPHM} phenomenological waveform \cite{IMRPhenomXPHM} implemented in the \texttt{PyCBC} \cite{PyCBC} package. 
    The duration of the waveform is set to be $\SI{16}{\second}$, the lower frequency $f_\text{low}=\SI{5}{\hertz}$ and the sampling rate $f_\text{s} = 1024\,\si{Hz}$.

    \item \textit{Projection and standardization:} 
    Then, to obtain a single batch $B$ of data for the epoch we randomly pick $K\times B$ templates from the bank. Then we draw an equivalent number of samples from the \textit{extrinsic} parameter prior (Tab. \ref{tab: prior distributions}) to individually project those waveforms into the ET channels before summing together the $K$ overlapping sources. For the projection we adopt the XYZ base instead of the AET one. Finally we generate Gaussian colored noise from the reference ASD \cite{ET_curve} and whiten data. 
    We repeat step (ii) operations for $N_\text{steps}$ times during every epoch.
\end{enumerate}

\noindent To speed-up data generation, each operation in step (ii) is vectorized through \texttt{PyTorch} \cite{paszke2017automatic} to leverage GPUs parallel computational power. Indeed, on a Dell PowerEdge R7425 machine equipped with a 64 core \textsc{AMD Epyc} CPU and two \textsc{NVIDIA A30} GPUs
step (i) takes $\sim30\si{s}$ while step (ii) is almost instantaneous. \\
The total amount of different representations seen by the network during a training is therefore given by
\begin{equation}\label{eq: number of samples}
    N = B \times N_\text{steps} \times E
\end{equation}
with $E$ the total number of epochs. 

This simulation strategy has different advantages. First, it allows the training dataset to be arbitrarily large, and the sampling of extrinsic parameters effectively acts as a form of data augmentation. Moreover, since new samples are produced at every epoch, the network is robust against overfitting, although the noise distribution remain consistent across all simulations.

\begin{table}[!t]
\centering
\begin{tabular}{ c c }
\hline
\hline
  parameter & prior  \\ %[1.5ex] 
\hline
\hline
\multicolumn{2}{c}{\textbf{Intrinsic}} \\
$m_{1}$ [$M_\odot$]     &      $\mathcal{U}(100, 800)$ \\
$m_2 \leq m_1$ [$M_\odot$]     &      $\mathcal{U}(100, 650)$   \\
inclination $\iota$   & $\sin (0, \pi)$\\
coalescence phase $\phi$   & $\mathcal{U}\,(0, 2\pi)$ \\
\hline
\multicolumn{2}{c}{\textbf{Extrinsic}} \\
SNR $\rho_\text{opt}$ & $\mathcal{U}(10, 150)$   \\
polarization $\psi$  & $\mathcal{U}\,(0, 2\pi)$   \\
right ascension $\alpha$  & $\mathcal{U}\,(0, 2\pi)$  \\
declination $\delta$  & $\cos\,(-\pi/2, \pi/2)$  \\
$t_\text{merger}$  & $\mathcal{U}(-\frac{2}{3}T_\text{obs}, 0)$  \\
$t_\text{GPS}$  & fixed $1399075278$  \\
\hline
\hline
\end{tabular}

\caption{Priors over the parameters for the simulated waveforms. Top half denotes the \textit{intrinsic parameters}, while the bottom half denotes the \textit{extrinsic} parameters. $T_\text{obs}$ is the duration of the observation, which in this work is 16 seconds.}
\label{tab: prior distributions}
\end{table}

In Tab. \ref{tab: prior distributions} we list the priors over the waveform's parameters. The range for the masses $m_{1,2}$ refers to the value in the detector frame, accounting for the cosmological redshift \cite{mass_redshift_degeneracy} (see Eq. \eqref{eq: redshift mass})
\begin{equation}\label{eq: redshift mass}
    m_{i}^{\text{det}} = (1+z)\,m_{i}^{\text{source}}
\end{equation}
We implicitly assumed a maximum redshift value $z_\text{max} = 10$ \cite{ET, ET_science_case_different_designs}, and focused on shorter duration \textit{high mass} waveforms.
\noindent When projecting, instead of rescaling the waveform amplitude in terms of the luminosity distance, we compute the \textit{optimal} signal-to-noise-ratio (SNR) as defined in Eq. \eqref{eq: SNR}, assuming the design ET-D sensitivity \cite{ET_curve} for $S_\text{n}(f)$

\begin{equation}\label{eq: SNR}
    \rho^2_\text{opt} = 4 \int_{0}^\infty df\; \frac{|\hat{h}(f)|^2}{S_\text{n}(f)} 
\end{equation}

\noindent then we rescale $h(t)$ to have a network SNR uniformly distributed (Tab. \ref{tab: prior distributions}) 

\begin{equation}\label{eq: SNR network}
    \rho_\text{net} = \sqrt{\sum_\text{detectors} \rho_\text{det}^2}
\end{equation}

In standard Bayesian inference, when sampling the parameter space $\{\vect{\theta}\}_k$ for a set of $K$ overlapped sources, there is typically symmetry under permutation of the $k$ indices. This means that there are multiple possible ways to assign the $k$ labels, which may result in degenerate posterior samples.
This is referred as the \textit{label-switching} problem which can be tackled involving specific prior sampling strategies and/or symmetrizations of the likelihood \cite{Buscicchio_label_switching}.
Working in a SBI, and in particular \textit{likelihood free}, approach we are instead forced to adopt a source ordering when simulating the training dataset.

\noindent The overlapped signal are therefore ordered according to their network SNR
\begin{equation}\label{eq: SNR requirement}
    \rho_\text{net}^i \geq \rho_\text{net}^j \qquad \text{if}\qquad i\leq j
\end{equation}

\begin{figure}
    \centering
    \hspace{-1cm} % Adjust the value as needed
    \includegraphics[width=0.98\linewidth]{ 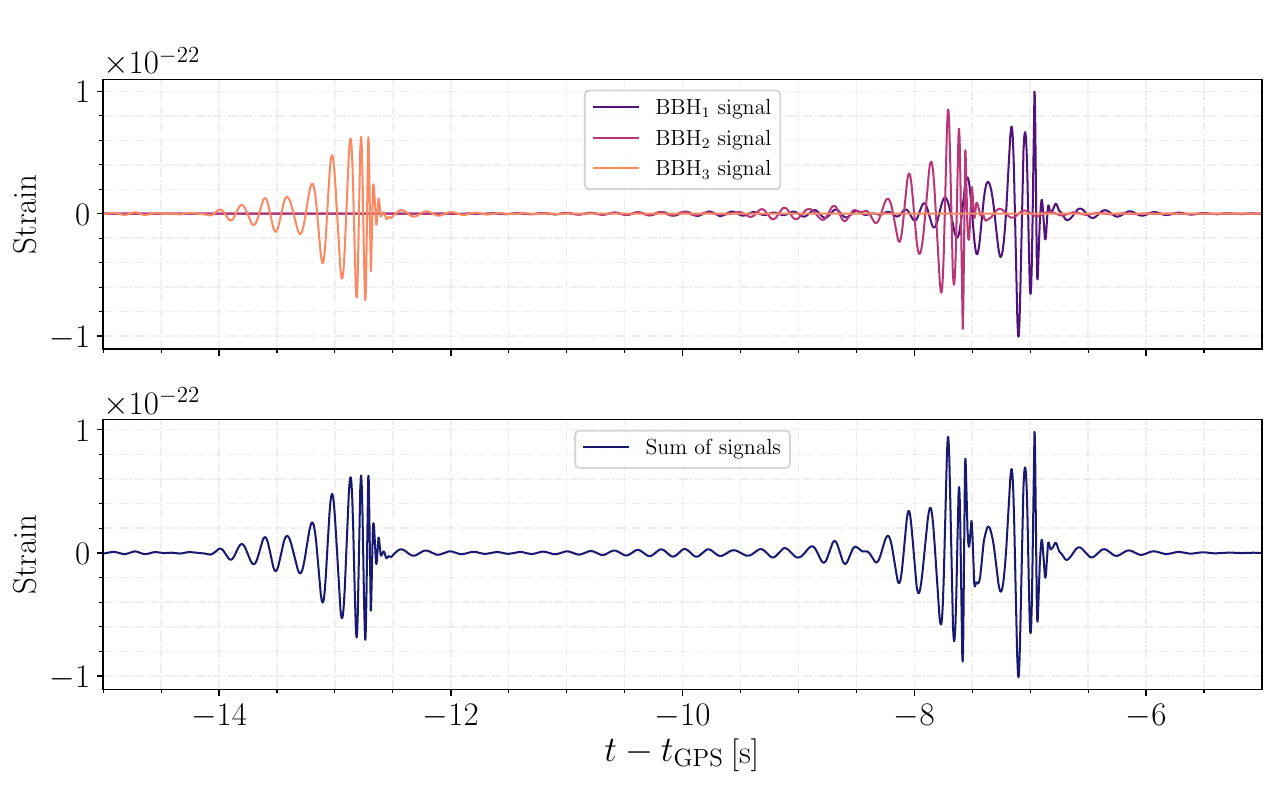}
    \caption{Noiseless injection from the training dataset.
    \textit{Top panel:} Three BBH signals projected into the E1 channel are displayed individually. \textit{Bottom panel:} The resulting summed waveform. }
    \label{fig:example}
\end{figure}

\noindent The introduction of this kind of hierarchy improved the recovering of the single sources. %Furthermore, it allows a more confident inference on louder signals which could potentially allow to perform model selection over the number of sources.
There still might be cases in which posterior samples are affected by mild degeneracy due signals with close $\rho_\text{net}$ or parameters as in the case of gravitational lenses \cite{GW_Lensing}. We further discuss this in Sec. \ref{sec: results and discussion} as well as a possible mitigation strategy.
%\verify{I would add also other reasons. e.g. having a more confident inference on louder signal helps when performing model selection over the number of sources or in the case of lensed signals \cite{GW_Lensing}}
Fig. \ref{fig:example} shows an example of a simulated noiseless injection. As from Tab. \ref{tab: prior distributions}, the merger time is randomly sampled within the time range of the sample, while making sure for the template to be fully inside the interval. %This makes the overlap of the signals completely random. \\

 \subsection{Training}\label{sec: training}

We have trained the network to estimate the following set of parameters: $\{ \vect{\theta}_\text{gw}\}_k = \{M_k\,, \mathcal{M}_k\,, q_k\,, t_{merger,\, k}\}$ with $k=1,2,3$ identifying the three injected binaries. $M$ is the total mass of the binary, $\mathcal{M}$ the chirp mass, $q$ the mass ratio and $t_{merger}$ the time of the coalescence.
For the training phase, we adopted a batch size $B = 128$ samples, with the model updated over $1000$ $N_\text{steps}$ per training epoch. Validation was instead conducted with 150 steps per epoch.

Training lasted for 300 epochs, to ensure model convergence, and took around 3 days on one NVIDIA A30 GPU. Optimization was carried out using the \textsc{ADAM} stochastic gradient optimizer \cite{kingma2017adam} with the loss function defined by Eq. \eqref{eq: KL loss}.

\begin{figure}
    \centering
    \hspace{-0.8cm}
    \includegraphics[width=\linewidth]{ 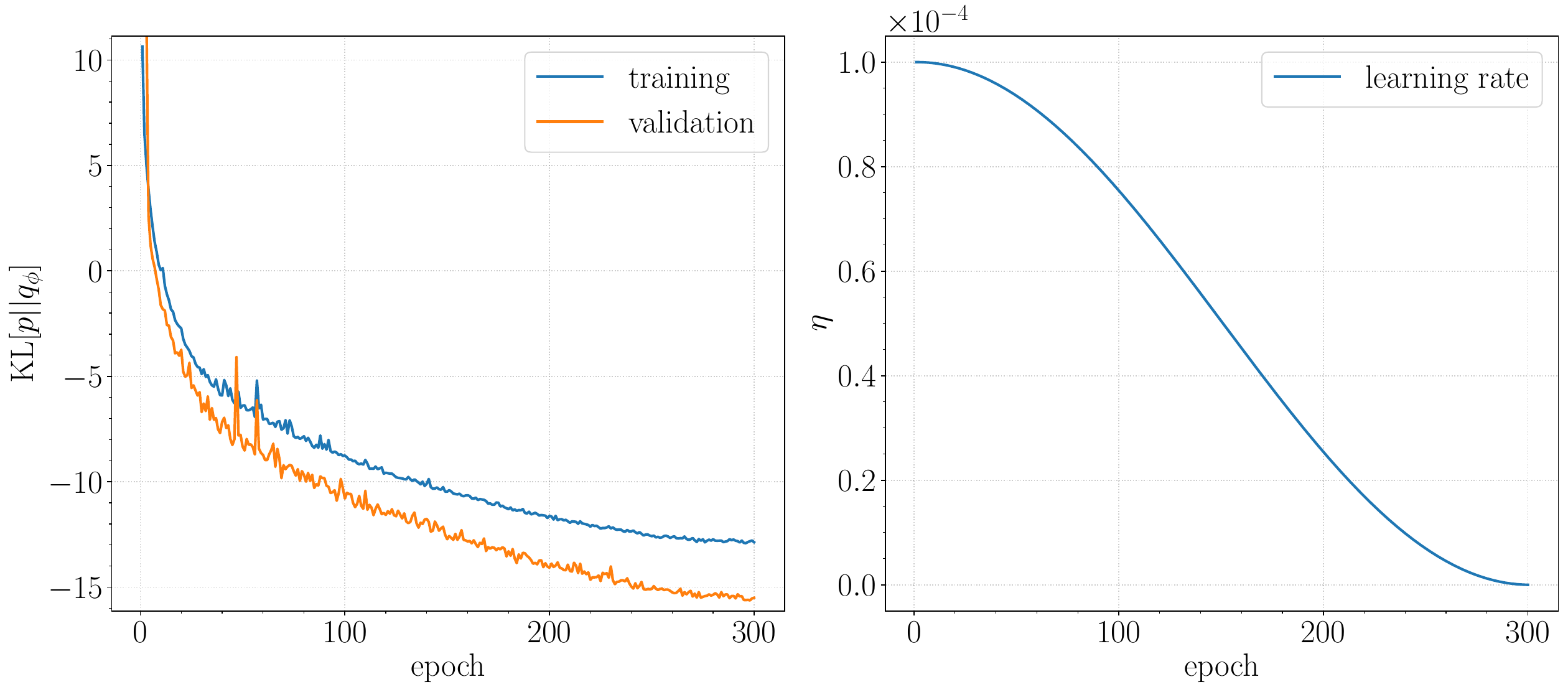}
    \caption{\textit{Left:} The evolution of the KL Divergence $\mathrm{KL}[p \| q_\phi]$ loss for both training (blue) and validation (orange) over 300 epochs. \textit{Right:} The learning rate $\eta$ decreases according to a \textit{cosine annealing} schedule.}
    \label{fig:training_curve}
\end{figure}

As determined by Eq. \eqref{eq: number of samples}, the number of samples processed by the network is $N = 3.84\times 10^7$. The learning curves are shown in Fig. \ref{fig:training_curve} where we plot the evolution of both training and validation loss values averaged over every single epoch batches. The validation curve lies below the training curve since the training loss is computed as the average over all $N_\text{steps}$ iterations, during which the model is continuously optimized, whereas the validation loss reflects the model’s performance at the final iteration and stabilizes around that value.

%-----------------------RESULTS----------------------------
\section{Results and discussion}\label{sec: results and discussion}

\begin{figure}
    \centering
    \includegraphics[width=\linewidth]{ 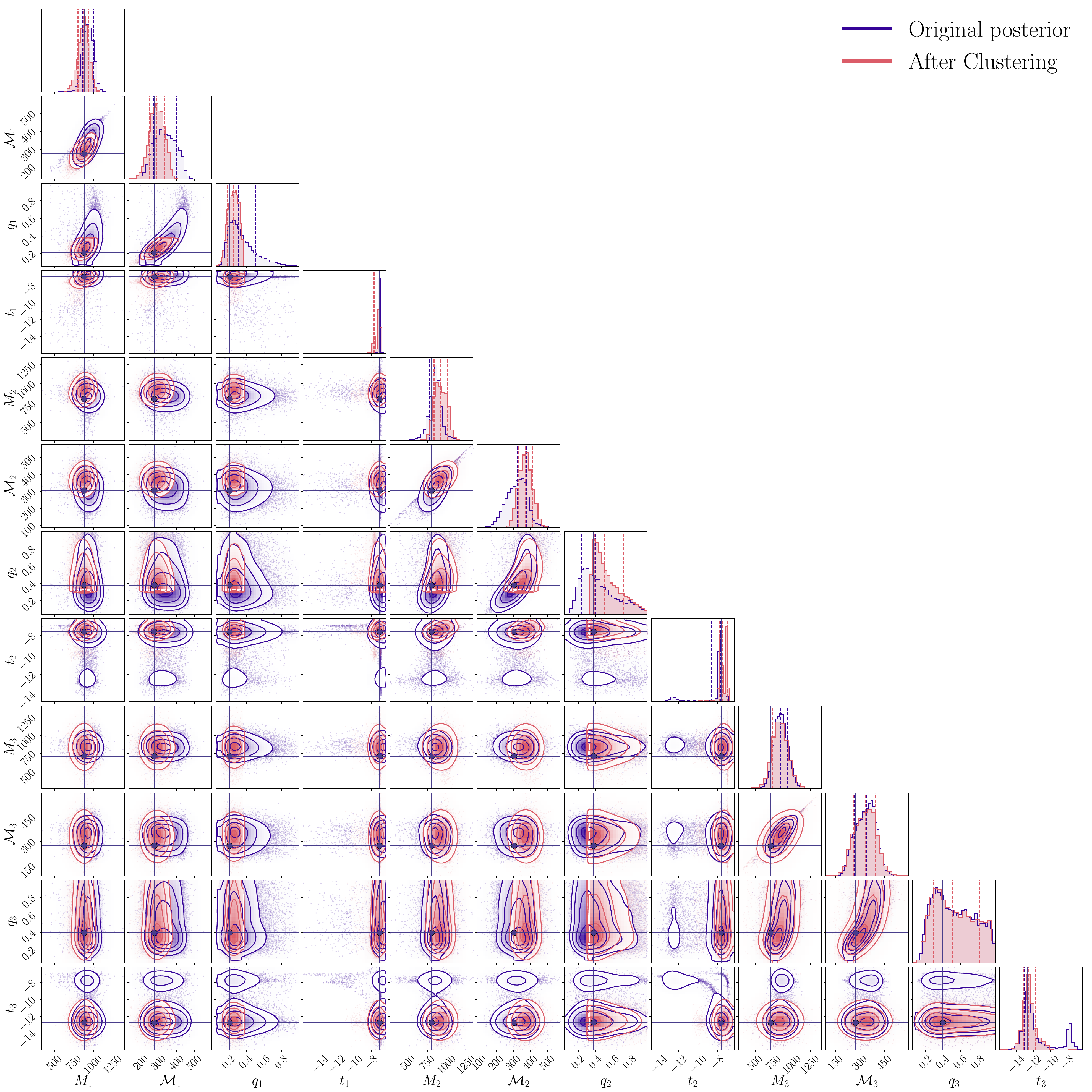}
    \caption{Inferred posterior distribution for the same injection from the test dataset of Fig. \ref{fig:example} with $\rho_\text{net}$ = 108,  87,  70.
    In blue the raw posterior samples drawn by the model, while in pink the same posterior samples after being processed with the clustering algorithm.
    Blue solid lines indicates instead the true injected parameters. The dotted lines indicate the 68\% credible interval around the median of the posterior distribution (central dotted line). We notice how the bimodality in $t_{{\text{merger}}, \,1}$ and $t_{{\text{merger}}, \,2}$ due to their relatively close $\rho_\text{net}$ gets correctly removed by the clustering strategy and how the posterior narrows around the true values. }
    \label{fig: corner plot}
\end{figure}

The inferred posterior for the test injection of Fig. \ref{fig:example} is shown in Fig. \ref{fig: corner plot}. Our algorithm takes $\sim1\si{s}$ to generate a full posterior with $10^4$ samples.  As it can be observed in the plot (the blue one in Fig. \ref{fig: corner plot}), our model correctly recovers the parameters of the three binaries. We notice however that the posterior exhibits multimodalities in some parameters due to the presence secondary peaks in the marginal distribution. %This is probably due to the network's difficulty in fully distinguishing the three different signals. 
To mitigate this effect we introduced a final post-processing of the posterior samples with the aim of reassigning the correct $k$ label to each posterior sample.
To do so, we exploited a clustering algorithm, in particular the \textit{Spectral Clustering} \cite{spectral_clustering, Gerosa_gwlabel}. The choice of this specific algorithm is motivated by the interesting findings of \cite{Gerosa_gwlabel}. % which is well suited when the distribution of cluster's points is highly non-convex. 
\noindent More in detail, the clustering happens over the eigenvectors of an \textit{affinity} matrix $A_{i, j}\geq 0$ which quantifies the similarity of two given samples $(x_i, x_j)$.
We construct the affinity matrix using $k$-Nearest Neighbors \cite{kNN} with $k_\text{neigh}=1000$ and $\mathbb{L}_1$ distance applied to the individual $\{\vect{\theta_\text{gw}}\}_k$ sets of posterior samples. For the Spectral Clustering we adopt the implementation of the \textsc{Scikit-Learn} library \cite{scikit_learn}.
Being an unsupervised algorithm, the newly clustered posterior samples might not in general reflect the original $\rho_\text{net}$ sorting. We therefore add a final step to map the new labels backward to the original ones by applying the Hungarian algorithm \cite{hungarian_algorithm} to minimize the difference between the posterior median values for each of the $(k^\text{old}, k^\text{new})$ set of cluster labels.
This phase of postprocessing takes $\sim 5\si{s} $ to complete.
From the corner plot in Fig. \ref{fig: corner plot} it is possible to see the effectiveness of this procedure. In particular, it can be seen that the bimodality in $t_\text{merger}$ is correctly removed and that almost all the marginal posteriors shrink and narrow around the true injected value.
\begin{figure}[!t]
    \centering
    \includegraphics[width=0.8\linewidth]{ 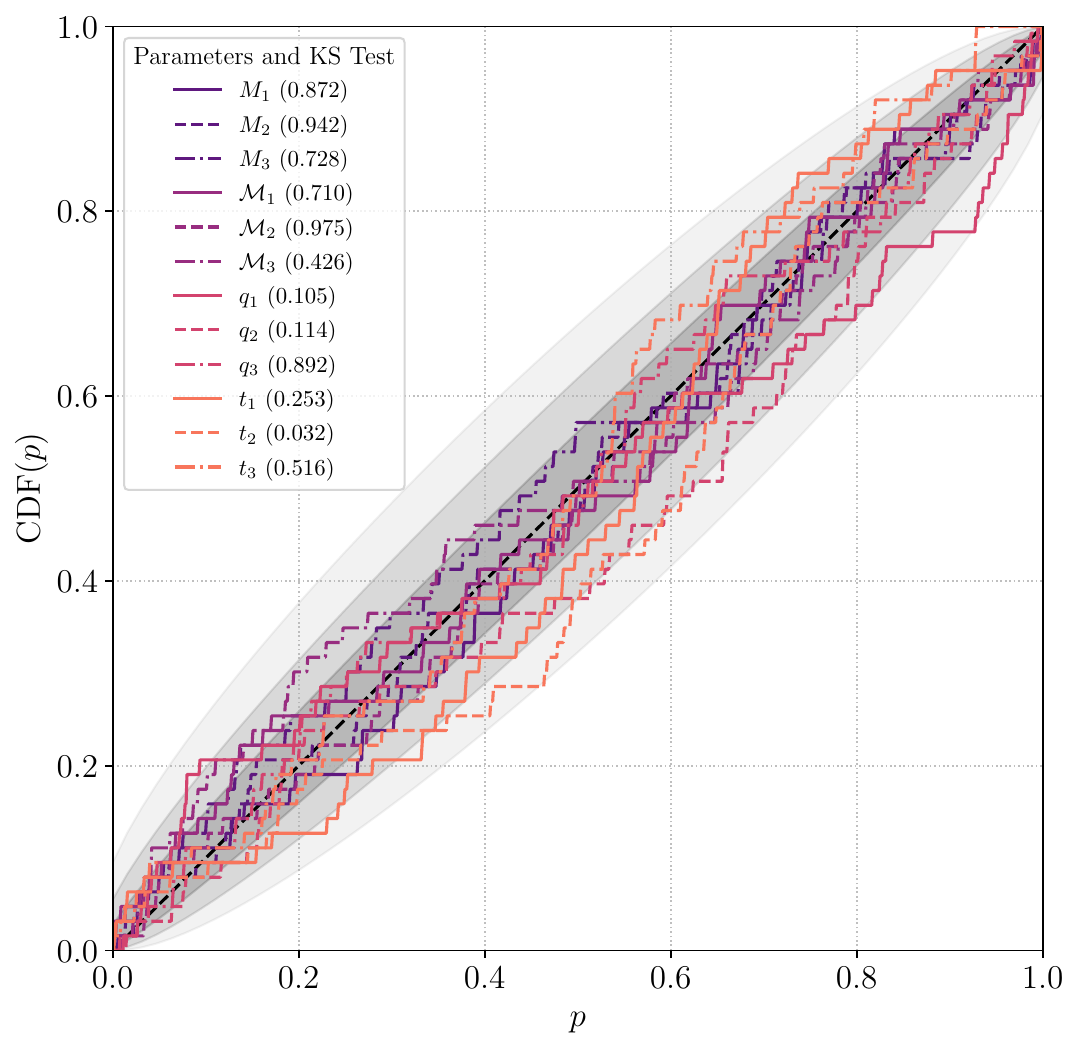}
    \caption{PP plot for the posterior distributions of 64 injections. For each of the inferred parameter we report also the $p$-values from the KS test. The combined $p$-value is $0.489$. }
    \label{fig: PP plot}
\end{figure}

To assess the reliability of the parameter estimation and statistically test our network, we realized a Probability-Probability plot (PP-plot) \cite{Gabbard_Vitamin, Veitch:2014wba} with posteriors drawn from 64 different test injections.
The PP-plot indeed compares the probability $p$ contained in credible intervals (x-axis) to the actual cumulative distribution of true values falling within those intervals (y-axis).
In a well-calibrated model, the points should align with the 45 degree diagonal, indicating that the predicted intervals accurately reflect the true values. In our case, shown in Fig. \ref{fig: PP plot}, most curves align closely with the diagonal, within the confidence bands, indicating that the model performs reasonably for these parameters. For each parameter we also performed a KS test whose combined $p$-value is $0.489$.

%To further assess the performance and robustness of our model we decided to test it under different levels of ``data difficulty", namely the level of correlation between the three injected BBH signals. 
To further assess the performance and robustness of our model we decided to test it under different correlation levels between the three BBH signals.
To do so, we determine how many of the three BBH signals are correlated with each other. The higher the number, the more challenging we expect the sample to be for the network. 
\noindent Specifically, we calculate the correlation between every individual signal $h_k ^\text{det}$ and the sum of the other two $\sum_{j\neq k} h_j^\text{det}$ in the following way 

\begin{subequations}
    \begin{align}
        \mathcal{C}_{k, \, \text{det}} &= \mathcal{P} \left[ h_k ^\text{det} \,\Big|  \sum_{j\neq k} h_j^\text{det} \right]\label{eq: single pearson}\\
        \mathcal{C}_k \;&= \frac{1}{\sqrt{N_\text{det}}} \left(\; \sum_{\text{detectors}} \mathcal{C}_{k,\,\text{det}}^2 \;\right)^{1/2}\label{eq: detector pearson statistic}
    \end{align}
\end{subequations}

\noindent where $\mathcal{P}$ is the \textit{Pearson Correlation coefficient} which, for two random variables $X$ and $Y$, takes the following form

\begin{equation}\label{eq: Pearson}
    \mathcal{P}[{X|Y}] = \frac{\text{cov}(X,Y)}{\sigma_X\sigma_Y} =\dfrac{\mathbb{E}[(X - \mu_X)(Y-\mu_Y)]}{\sigma_X\sigma_Y}
\end{equation}

\noindent with the term $\sigma_{X,Y}$ representing the standard deviation and $\text{cov}(X,Y)$ the covariance of $X, Y$.
Our correlation statistic is then defined as
\begin{equation}\label{eq: pearson statistic}
    \mathcal{C} = \sum_k \Theta(C_k - 0.05)
\end{equation}
with $\Theta(x)$ being the Heaviside step function.
\noindent Given the definition of Eq. \eqref{eq: pearson statistic}, $\mathcal{C}$ can assume discrete values $[0, \cdots, K]$ which represents the number of BBH signals whose correlation with at least one other signal
exceeds a 5\% threshold. Therefore $\mathcal{C}=0$ ($\mathcal{C}=K$) represent uncorrelated (highly correlated) injections.

\begin{figure}[!t]
    \centering
    \hspace{-1cm}
    \includegraphics[scale=0.35]{ 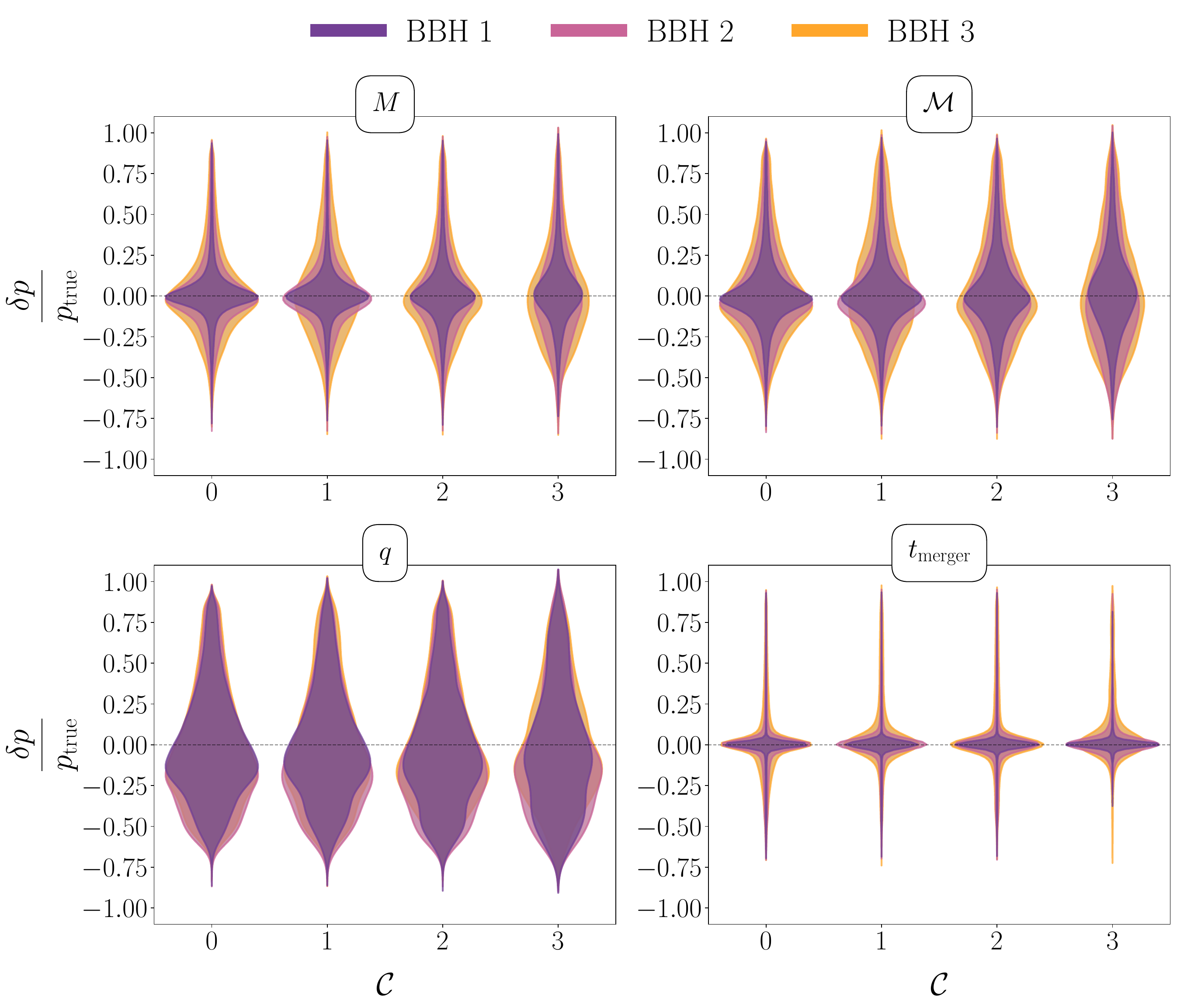}
    \caption{Violin plots showing the relative error $\delta p/ p_{\text{true}}$ distributions for the four parameters estimated in our analysis: total mass $M$, chirp mass $\mathcal{M}$, mass ratio $q$ and merger time $t_\text{merger}$. The four classes on the x-axis $\{0, 1, 2, 3\}$, represent the metric defined in Eq. \eqref{eq: pearson statistic}, indicating how many signals exhibit a correlation $C_k>0.05$. The different colors correspond the three injected signals. %purple for $\text{BBH}_1$, blue for $\text{BBH}_2$, yellow for $\text{BBH}_3$. 
    }
    \label{fig: pearson violin plot}
\end{figure}

We then ran again the testing procedure with $10^5$ simulations and compared the distributions of the relative errors $\delta p/p_\text{true}$ for the posterior medians with respect to the true values. We show the result in Fig. \ref{fig: pearson violin plot}. As it can be observed, the network performs well and exhibits the same behavior regardless the correlation. A slight broadening of the relative error distribution, as expected, can be seen as the samples become more difficult, i.e., when more signals are correlated. Moreover, the distributions exhibit an expected behavior related to the $\rho_\text{net}$ sorting of the BBHs as the error distribution broadens for fainter signals.
We notice that among the parameters, the mass ratio's violin plots have a higher spread and a slight off-zero peak. This might originate from eventual non-gaussian $q$ posteriors, as seen e.g. in the one of Fig. \ref{fig: corner plot}. %and is not in general an indicator of a biased PE, given that $\delta p / p_\text{true}$ does not take into account the marginal posterior's width. 
%We recall as well that in this analysis the SNR of the injection wasn't fixed but instead followed the training prior.

%-----------------------CONCLUSION-------------------------
\newpage
\section{Conclusions}\label{sec: conclusion}
In this work, we propose a novel strategy to perform Bayesian parameter estimation for multiple overlapped signals in the Einstein Telescope data based on Transformers and Normalizing Flows. 
In fact, inference on single source parameters in next-generation detectors is going to be significantly impacted by the inevitable overlap of multiple sources due to a higher detection rate $\mathcal{O}(10^5 \text{ ev/yr})$ \cite{Iacovelli:2022bbs} and the longer duration of the signals \cite{ET_blue_book}.
The Transformer architecture that we implemented efficiently encodes the complex gravitational wave data. We used it in synergy with a Normalizing Flow model to efficiently estimate posterior distributions for three overlapping BBH signals. We tested our model on a statistically relevant number of samples to show how it successfully recovers the parameters of the sources. Additionally, we leveraged the Spectral Clustering algorithm \cite{Gerosa_gwlabel} to remove spurious degeneracies in the inferred posteriors.\\
Furthermore we tested our method against different correlation levels among the signals. Our results exhibit no relevant dependencies, thus indicating that the model maintains the same level of accuracy regardless of the difficulty of data. 
An important aspect to highlight is the small inference time ($\sim 1\si{s}$).
All these features make this architecture well-suited for the low-latency analysis of candidate gravitational wave events, thereby enabling rapid electromagnetic follow-ups, which are critical for multimessenger astronomy observations. 
While our results highlight the potential of Transformer-based architectures in gravitational-wave inference, we believe that a significant margin for improvement remains. Some that we intend to explore in future works include extending the inference parameter space, with particular attention to sky localization and spins, as well as an increased and eventually varying number of overlapping mergers. This might require an improvement of the Convolutional Embedding component of the network.  
The approach presented here is flexible and scalable enough to be easily adapted to include BNS signals. Therefore, the promising results obtained in this work represent a viable pathway for the development of next-generation data analysis pipelines suitable for future and current gravitational wave detectors.

\section*{Code Availability}
Code for \textsc{KENN} is publicly available at ref. \cite{KENN_code}, while \textsc{HYPERION} at ref. \cite{HYPERION_code}.

\section*{Acknowledgments}
The authors gratefully acknowledge Davide Gerosa, Rodrigo Tenorio, Matteo Della Rocca, Ilaria Caporali and Chris Messenger for valuable discussions, and NVIDIA Corporation for the donation of the two A30 GPUs used in this work.
L.P. acknowledges the support of the PhD scholarship in Physics ``High Performance Computing and Innovative Data Analysis Methods in Science" (Cycle XXXVIII, Ministerial Decree no. 351/2022) and received funding from the European Union Next-Generation EU - National Recovery and Resilience Plan (NRRP) – MISSION 4 COMPONENT 1, INVESTMENT N.4.1 – CUP N.I51J22000630007. \\
F.D.S. is supported by ERC Starting Grant No.~945155--GWmining, Cariplo Foundation Grant No.~2021-0555, MUR PRIN Grant No.~2022-Z9X4XS, MUR Grant ``Progetto Dipartimenti di Eccellenza 2023-2027'' (BiCoQ),
and the ICSC National Research Centre funded by NextGenerationEU. I.S.H was supported by the Science and Technology Facilities Council
(STFC) grants ST/V001736/1 and ST/V005634/1. This manuscript reflects only the authors’ views and opinions, neither the European Union nor the European Commission can be considered responsible for them. 

%bibliography
\section*{References}
% Load IOP bibliography style and AAS macros
\bibliographystyle{custom-iopart-num}
\bibliography{bibliography}

\providecommand{\newblock}{}
\begin{thebibliography}{10}
\expandafter\ifx\csname url\endcsname\relax
  \def\url#1{{\tt #1}}\fi
\expandafter\ifx\csname urlprefix\endcsname\relax\def\urlprefix{URL }\fi
\providecommand{\eprint}[2][]{\href{http://arxiv.org/abs/#2}{arXiv:#2}}

\bibitem{GW150914}
Abbott B~P {\em et~al.\/} (LIGO Scientific, Virgo) 2016 {\em {Observation of Gravitational Waves from a Binary Black Hole Merger}\/} \href{http://dx.doi.org/10.1103/PhysRevLett.116.061102}{ {\em Phys. Rev. Lett.\/} {\bf 116} 061102 } [\eprint{1602.03837}]

\bibitem{cumulative_detection_rate}
Broekgaarden F~S, Banagiri S and Payne E 2024 {\em {Visualizing the Number of Existing and Future Gravitational-wave Detections from Merging Double Compact Objects}\/} \href{http://dx.doi.org/10.3847/1538-4357/ad4709}{ {\em Astrophys. J.\/} {\bf 969} 108 } [\eprint{2303.17628}]

\bibitem{ET}
Maggiore M {\em et~al.\/} (ET) 2020 {\em {Science Case for the Einstein Telescope}\/} \href{http://dx.doi.org/10.1088/1475-7516/2020/03/050}{ {\em JCAP\/} {\bf 03} 050 } [\eprint{1912.02622}]

\bibitem{Cosmic_Explorer}
Evans M {\em et~al.\/} 2021 {\em {A Horizon Study for Cosmic Explorer: Science, Observatories, and Community}\/} [\eprint{2109.09882}]

\bibitem{LISA}
{Amaro-Seoane} P {\em et~al.\/} 2017 {\em Laser interferometer space antenna\/} [\eprint{1702.00786}]

\bibitem{Iacovelli:2022bbs}
Iacovelli F {\em et~al.\/} 2022 {\em {Forecasting the Detection Capabilities of Third-generation Gravitational-wave Detectors Using GWFAST}\/} \href{http://dx.doi.org/10.3847/1538-4357/ac9cd4}{ {\em Astrophys. J.\/} {\bf 941} 208 } [\eprint{2207.02771}]

\bibitem{ET_science_case_different_designs}
Branchesi M {\em et~al.\/} 2023 {\em {Science with the Einstein Telescope: a comparison of different designs}\/} \href{http://dx.doi.org/10.1088/1475-7516/2023/07/068}{ {\em JCAP\/} {\bf 07} 068 } [\eprint{2303.15923}]

\bibitem{PE_challenges_3G}
Baker A~M {\em et~al.\/} 2025 {\em {Significant challenges for astrophysical inference with next-generation gravitational-wave observatories}\/} [\eprint{2503.04073}]

\bibitem{Pizzati_et_al_overlap}
Pizzati E {\em et~al.\/} 2022 {\em {Toward inference of overlapping gravitational-wave signals}\/} \href{http://dx.doi.org/10.1103/PhysRevD.105.104016}{ {\em Phys. Rev. D\/} {\bf 105} 104016 } [\eprint{2102.07692}]

\bibitem{Overlapping_samajdar}
Samajdar A {\em et~al.\/} 2021 {\em {Biases in parameter estimation from overlapping gravitational-wave signals in the third-generation detector era}\/} \href{http://dx.doi.org/10.1103/PhysRevD.104.044003}{ {\em Phys. Rev. D\/} {\bf 104} 044003 } [\eprint{2102.07544}]

\bibitem{impact_overlapping_GR_tests}
Dang Y, Wang Z, Liang D and Shao L 2024 {\em {Impact of Overlapping Signals on Parameterized Post-Newtonian Coefficients in Tests of Gravity}\/} \href{http://dx.doi.org/10.3847/1538-4357/ad2e00}{ {\em Astrophys. J.\/} {\bf 964} 194 } [\eprint{2311.16184}]

\bibitem{cbc_subtraction_for_sgwb_2}
Sachdev S, Regimbau T and Sathyaprakash B~S 2020 {\em {Subtracting compact binary foreground sources to reveal primordial gravitational-wave backgrounds}\/} \href{http://dx.doi.org/10.1103/PhysRevD.102.024051}{ {\em Phys. Rev. D\/} {\bf 102} 024051 } [\eprint{2002.05365}]

\bibitem{cbc_subtraction_for_sgwb}
Zhou B {\em et~al.\/} 2023 {\em {Subtracting compact binary foregrounds to search for subdominant gravitational-wave backgrounds in next-generation ground-based observatories}\/} \href{http://dx.doi.org/10.1103/PhysRevD.108.064040}{ {\em Phys. Rev. D\/} {\bf 108} 064040 } [\eprint{2209.01310}]

\bibitem{overlapping_cosmological_sgwb}
Himemoto Y, Nishizawa A and Taruya A 2021 {\em Impacts of overlapping gravitational-wave signals on the parameter estimation: Toward the search for cosmological backgrounds\/} \href{http://dx.doi.org/10.1103/PhysRevD.104.044010}{ {\em Phys. Rev. D\/} {\bf 104}(4) 044010 }

\bibitem{notching_cosmological_sgwb}
Zhong H {\em et~al.\/} 2024 {\em {Searching for cosmological stochastic backgrounds by notching out resolvable compact binary foregrounds with next-generation gravitational-wave detectors}\/} \href{http://dx.doi.org/10.1103/PhysRevD.110.064047}{ {\em Phys. Rev. D\/} {\bf 110} 064047 } [\eprint{2406.10757}]

\bibitem{ET_blue_book}
Abac A {\em et~al.\/} 2025 {\em {The Science of the Einstein Telescope}\/} [\eprint{2503.12263}]

\bibitem{Cuoco:2024cdk}
Cuoco E {\em et~al.\/} 2025 {\em {Applications of machine learning in gravitational-wave research with current interferometric detectors}\/} \href{http://dx.doi.org/10.1007/s41114-024-00055-8}{ {\em Living Rev. Rel.\/} {\bf 28} 2 } [\eprint{2412.15046}]

\bibitem{ML_physics_GW}
Agarwal M {\em et~al.\/} 2023 [\eprint{2306.08106}]

\bibitem{Nousi_AresGW}
Nousi P {\em et~al.\/} 2023 {\em {Deep residual networks for gravitational wave detection}\/} \href{http://dx.doi.org/10.1103/PhysRevD.108.024022}{ {\em Phys. Rev. D\/} {\bf 108} 024022 } [\eprint{2211.01520}]

\bibitem{Bini_ML_cWB}
Bini S {\em et~al.\/} 2023 {\em {An autoencoder neural network integrated into gravitational-wave burst searches to improve the rejection of noise transients}\/} \href{http://dx.doi.org/10.1088/1361-6382/acd981}{ {\em Class. Quant. Grav.\/} {\bf 40} 135008 } [\eprint{2303.05986}]

\bibitem{ML_detection_SNe}
Chan M~L, Heng I~S and Messenger C 2020 {\em {Detection and classification of supernova gravitational wave signals: A deep learning approach}\/} \href{http://dx.doi.org/10.1103/PhysRevD.102.043022}{ {\em Phys. Rev. D\/} {\bf 102} 043022 } [\eprint{1912.13517}]

\bibitem{ML_detection_long_transients}
Modafferi L~M, Tenorio R and Keitel D 2023 {\em {Convolutional neural network search for long-duration transient gravitational waves from glitching pulsars}\/} \href{http://dx.doi.org/10.1103/PhysRevD.108.023005}{ {\em Phys. Rev. D\/} {\bf 108} 023005 } [\eprint{2303.16720}]

\bibitem{Dingo}
{Dax} M {\em et~al.\/} 2021 {\em {Real-Time Gravitational Wave Science with Neural Posterior Estimation}\/} \href{http://dx.doi.org/10.1103/PhysRevLett.127.241103}{ {\em Phys. Rev. Lett.\/} {\bf 127} 241103 } [\eprint{2106.12594}]

\bibitem{Nessai}
Williams M~J, Veitch J and Messenger C 2021 {\em {Nested sampling with normalizing flows for gravitational-wave inference}\/} \href{http://dx.doi.org/10.1103/PhysRevD.103.103006}{ {\em Phys. Rev. D\/} {\bf 103} 103006 } [\eprint{2102.11056}]

\bibitem{Gabbard_Vitamin}
Gabbard H {\em et~al.\/} 2022 {\em {Bayesian parameter estimation using conditional variational autoencoders for gravitational-wave astronomy}\/} \href{http://dx.doi.org/10.1038/s41567-021-01425-7}{ {\em Nature Phys.\/} {\bf 18} 112--117 } [\eprint{1909.06296}]

\bibitem{Razzano_gwitchhunters}
Razzano M {\em et~al.\/} 2023 {\em {GWitchHunters: Machine learning and citizen science to improve the performance of gravitational wave detector}\/} \href{http://dx.doi.org/10.1016/j.nima.2022.167959}{ {\em Nucl. Instrum. Meth. A\/} {\bf 1048} 167959 } [\eprint{2301.05112}]

\bibitem{Gravity_Spy}
Zevin M {\em et~al.\/} 2017 {\em {Gravity Spy: Integrating Advanced LIGO Detector Characterization, Machine Learning, and Citizen Science}\/} \href{http://dx.doi.org/10.1088/1361-6382/aa5cea}{ {\em Class. Quant. Grav.\/} {\bf 34} 064003 } [\eprint{1611.04596}]

\bibitem{denoising_ML}
Bacon P, Trovato A and Bejger M 2023 {\em {Denoising gravitational-wave signals from binary black holes with a dilated convolutional autoencoder}\/} \href{http://dx.doi.org/10.1088/2632-2153/acd90f}{ {\em Mach. Learn. Sci. Tech.\/} {\bf 4} 035024 } [\eprint{2205.13513}]

\bibitem{Interferometer_Control_ML}
Ma P~X and Vajente G 2024 {\em {A deep learning technique to control the non-linear dynamics of a gravitational-wave interferometer}\/} \href{http://dx.doi.org/10.1088/1361-6382/ad1daa}{ {\em Class. Quant. Grav.\/} {\bf 41} 045003 } [\eprint{2302.07921}]

\bibitem{DeSanti_CE_HYPERION}
De~Santi F {\em et~al.\/} 2024 {\em {Deep learning to detect gravitational waves from binary close encounters: Fast parameter estimation using normalizing flows}\/} \href{http://dx.doi.org/10.1103/PhysRevD.109.102004}{ {\em Phys. Rev. D\/} {\bf 109} 102004 } [\eprint{2404.12028}]

\bibitem{vaswani2017attention}
Vaswani A {\em et~al.\/}{\em  2017 \/} [\eprint{1706.03762}]

\bibitem{kenton2019bert}
Devlin J {\em et~al.\/}{\em  2019 \/} [\eprint{1810.04805}]

\bibitem{achiam2023gpt}
{OpenAI} {\em et~al.\/} 2024 {\em {GPT-4 Technical Report}\/} [\eprint{2303.08774}]

\bibitem{radford2023robust}
Radford A {\em et~al.\/} 2022 {\em Robust speech recognition via large-scale weak supervision\/} [\eprint{2212.04356}]

\bibitem{LIGO_A+_curve}
Publicly available at \urlprefix\url{https://dcc.ligo.org/LIGO-T2000012/public}

\bibitem{ET_curve}
Publicly available at \urlprefix\url{https://apps.et-gw.eu/tds/?r=14065}

\bibitem{gw170817}
Abbott B~P {\em et~al.\/} (LIGO Scientific, Virgo) 2017 {\em {GW170817: Observation of Gravitational Waves from a Binary Neutron Star Inspiral}\/} \href{http://dx.doi.org/10.1103/PhysRevLett.119.161101}{ {\em Phys. Rev. Lett.\/} {\bf 119} 161101 } [\eprint{1710.05832}]

\bibitem{Janquart_joint_PE}
{Janquart} J {\em et~al.\/} 2023 {\em {Analyses of overlapping gravitational wave signals using hierarchical subtraction and joint parameter estimation}\/} \href{http://dx.doi.org/10.1093/mnras/stad1542}{ {\em Mon. Not. Roy. Astron. Soc.\/} {\bf 523} 1699--1710 } [\eprint{2211.01304}]

\bibitem{PE_overlapping_signals_2G}
Relton P and Raymond V 2021 {\em {Parameter estimation bias from overlapping binary black hole events in second generation interferometers}\/} \href{http://dx.doi.org/10.1103/PhysRevD.104.084039}{ {\em Phys. Rev. D\/} {\bf 104} 084039 } [\eprint{2103.16225}]

\bibitem{Antonelli_biases_overlapping}
Antonelli A, Burke O and Gair J~R 2021 {\em {Noisy neighbours: inference biases from overlapping gravitational-wave signals}\/} \href{http://dx.doi.org/10.1093/mnras/stab2358}{ {\em Mon. Not. Roy. Astron. Soc.\/} {\bf 507} 5069--5086 } [\eprint{2104.01897}]

\bibitem{Himemoto_overlapping_cgwb}
Himemoto Y, Nishizawa A and Taruya A 2021 {\em {Impacts of overlapping gravitational-wave signals on the parameter estimation: Toward the search for cosmological backgrounds}\/} \href{http://dx.doi.org/10.1103/PhysRevD.104.044010}{ {\em Phys. Rev. D\/} {\bf 104} 044010 } [\eprint{2103.14816}]

\bibitem{Wang_PE_bias_analusis_with_FM}
Wang Z {\em et~al.\/} 2024 {\em {Anatomy of parameter-estimation biases in overlapping gravitational-wave signals}\/} \href{http://dx.doi.org/10.1088/1361-6382/ad210b}{ {\em Class. Quant. Grav.\/} {\bf 41} 055011 } [\eprint{2304.06734}]

\bibitem{Source_confusion_BNS_minimal_overlapping_PE}
Johnson A~D, Chatziioannou K and Farr W~M 2024 {\em {Source confusion from neutron star binaries in ground-based gravitational wave detectors is minimal}\/} \href{http://dx.doi.org/10.1103/PhysRevD.109.084015}{ {\em Phys. Rev. D\/} {\bf 109} 084015 } [\eprint{2402.06836}]

\bibitem{merger_rate_density_used_in_samajdar_1}
{Belczynski} K {\em et~al.\/} 2017 {\em {On the likelihood of detecting gravitational waves from Population III compact object binaries}\/} \href{http://dx.doi.org/10.1093/mnras/stx1759}{ {\em Mon. Not. Roy. Astron. Soc.\/} {\bf 471} 4702--4721 } [\eprint{1612.01524}]

\bibitem{merger_rate_density_used_in_samajdar_2}
{Oguri} M 2018 {\em {Effect of gravitational lensing on the distribution of gravitational waves from distant binary black hole mergers}\/} \href{http://dx.doi.org/10.1093/mnras/sty2145}{ {\em Mon. Not. Roy. Astron. Soc.\/} {\bf 480} 3842--3855 } [\eprint{1807.02584}]

\bibitem{gwtc3_population}
Abbott R {\em et~al.\/} (KAGRA, VIRGO, LIGO Scientific) 2023 {\em {Population of Merging Compact Binaries Inferred Using Gravitational Waves through GWTC-3}\/} \href{http://dx.doi.org/10.1103/PhysRevX.13.011048}{ {\em Phys. Rev. X\/} {\bf 13} 011048 } [\eprint{2111.03634}]

\bibitem{cho2014learning}
Cho K {\em et~al.\/} 2014 {\em Learning phrase representations using rnn encoder-decoder for statistical machine translation\/} [\eprint{1406.1078}]

\bibitem{hochreiter1997long}
Hochreiter S and Schmidhuber J 1997 {\em Long short-term memory\/} \href{http://dx.doi.org/10.1162/neco.1997.9.8.1735}{ {\em Neural computation\/} {\bf 9} 1735--1780 }

\bibitem{dey2017gate}
Dey R and Salem F~M 2017 {\em Gate-variants of gated recurrent unit (gru) neural networks\/}

\bibitem{hochreiter2001gradient}
Informatik F, Bengio Y, Frasconi P and Schmidhuber J 2003 {\em Gradient flow in recurrent nets: the difficulty of learning long-term dependencies\/} {\em A Field Guide to Dynamical Recurrent Neural Networks. IEEE Press In\/}  237--243

\bibitem{DeepGenerativeModels}
Ruthotto L and Haber E 2021 {\em An introduction to deep generative modeling\/} [\eprint{2103.05180}]

\bibitem{papamakarios_flows}
Papamakarios G {\em et~al.\/} 2021 {\em Normalizing flows for probabilistic modeling and inference\/} [\eprint{1912.02762}]

\bibitem{RealNVP}
Dinh L, Sohl-Dickstein J and Bengio S 2017 {\em Density estimation using real nvp\/} [\eprint{1605.08803}]

\bibitem{dingo_15D_toy}
Green S~R and Gair J 2021 {\em {Complete parameter inference for GW150914 using deep learning}\/} \href{http://dx.doi.org/10.1088/2632-2153/abfaed}{ {\em Mach. Learn. Sci. Tech.\/} {\bf 2} 03LT01 } [\eprint{2008.03312}]

\bibitem{Naticchioni:sos_enattos}
Naticchioni L {\em et~al.\/} 2024 {\em {Characterizing the Sardinia candidate site for the Einstein Telescope}\/} \href{http://dx.doi.org/10.22323/1.441.0110}{ {\em Proceedings of Science\/} (TAUP2023) 110 }

\bibitem{IMRPhenomXPHM}
Pratten G {\em et~al.\/} 2021 {\em {Computationally efficient models for the dominant and subdominant harmonic modes of precessing binary black holes}\/} \href{http://dx.doi.org/10.1103/PhysRevD.103.104056}{ {\em Phys. Rev. D\/} {\bf 103} 104056 } [\eprint{2004.06503}]

\bibitem{PyCBC}
Nitz A {\em et~al.\/} 2024 {\em \href{https://doi.org/10.5281/zenodo.10473621}{gwastro/pycbc: v2.3.3} release of {PyCBC}\/}

\bibitem{paszke2017automatic}
{Paszke} A {\em et~al.\/} 2019 {\em {PyTorch: An Imperative Style, High-Performance Deep Learning Library}\/} [\eprint{1912.01703}]

\bibitem{mass_redshift_degeneracy}
Chen X, Li S and Cao Z 2019 {\em {Mass\textendash{}redshift degeneracy for the gravitational-wave sources in the vicinity of supermassive black holes}\/} \href{http://dx.doi.org/10.1093/mnrasl/slz046}{ {\em Mon. Not. Roy. Astron. Soc.\/} {\bf 485} L141--L145 } [\eprint{1703.10543}]

\bibitem{Buscicchio_label_switching}
Buscicchio R {\em et~al.\/} 2019 {\em Label switching problem in bayesian analysis for gravitational wave astronomy\/} \href{http://dx.doi.org/10.1103/PhysRevD.100.084041}{ {\em Phys. Rev. D\/} {\bf 100}(8) 084041 } [\eprint{1907.11631}]

\bibitem{GW_Lensing}
Poon J~S~C {\em et~al.\/} 2024 {\em {Galaxy lens reconstruction based on strongly lensed gravitational waves: similarity transformation degeneracy and mass-sheet degeneracy}\/} \href{http://dx.doi.org/10.1093/mnras/stae2660}{ {\em Mon. Not. Roy. Astron. Soc.\/} {\bf 536} 2212--2233 } [\eprint{2406.06463}]

\bibitem{kingma2017adam}
Kingma D~P and Ba J 2017 {\em Adam: A method for stochastic optimization\/} [\eprint{1412.6980}]

\bibitem{spectral_clustering}
{von Luxburg} U 2007 {\em {A Tutorial on Spectral Clustering}\/} \href{http://dx.doi.org/10.48550/arXiv.0711.0189}{ } [\eprint{0711.0189}]

\bibitem{Gerosa_gwlabel}
Gerosa D {\em et~al.\/} 2025 {\em {Which Is Which? Identification of the Two Compact Objects in Gravitational-Wave Binaries}\/} \href{http://dx.doi.org/10.1103/PhysRevLett.134.121402}{ {\em Phys. Rev. Lett.\/} {\bf 134} 121402 } [\eprint{2409.07519}]

\bibitem{kNN}
Goldberger J, Hinton G~E, Roweis S and Salakhutdinov R~R 2004 Neighbourhood components analysis {\em Advances in Neural Information Processing Systems\/} vol~17 (MIT Press) pp 513--520

\bibitem{scikit_learn}
{Pedregosa} F {\em et~al.\/} 2011 {\em {Scikit-learn: Machine Learning in Python}\/} \href{http://dx.doi.org/10.48550/arXiv.1201.0490}{ {\em Journal of Machine Learning Research\/} {\bf 12} 2825--2830 } [\eprint{1201.0490}]

\bibitem{hungarian_algorithm}
Crouse D~F 2016 {\em On implementing 2d rectangular assignment algorithms\/} \href{http://dx.doi.org/10.1109/TAES.2016.140952}{ {\em IEEE Transactions on Aerospace and Electronic Systems\/} {\bf 52} 1679--1696 }

\bibitem{Veitch:2014wba}
Veitch J {\em et~al.\/} 2015 {\em {Parameter estimation for compact binaries with ground-based gravitational-wave observations using the LALInference software library}\/} \href{http://dx.doi.org/10.1103/PhysRevD.91.042003}{ {\em Phys. Rev. D\/} {\bf 91} 042003 } [\eprint{1409.7215}]

\bibitem{KENN_code}
{Papalini} L 2025 {\em \textsc{KENN}\/} \url{https://github.com/luciapapalini/kenn-gw-transformer.git}

\bibitem{HYPERION_code}
{De Santi} F 2024 {\em \textsc{HYPERION}\/} \url{https://github.com/fdesanti/HYPERION}

\end{thebibliography}

\end{document}